\documentclass{ptapap}
\usepackage{graphicx}
\usepackage{hyperref}
\author{T. P. Adhikari}[CAMK]
\author{A. Rozanska}[CAMK]
\author{M. Sobolewska}[CAMK]
\author{B. Czerny}[CTP]
\affil[CAMK]{Nicolaus Copernicus Astronomical Center, Polish Academy of Sciences\\
  Bartycka 18, 00--716 Warszawa, Poland}
\affil[CTP]{Centre for Theoretical Physics, Polish Academy of Sciences\\
  Al. Lotnikow 32/46 02-668 Warszawa, Poland}

\title{On the warm absorber in AGN outflow}
\begin{document}
\maketitle
\begin{abstract}
Warm absorber (WA) is an ionised gas present in the line of sight to the AGN central engine.
The effect of the absorber is imprinted in the absorption lines observed in 
X-ray spectra of AGN. In this work, we model the WA in
Seyfert 1 galaxy Mrk 509 using its recently published shape of broad band spectral energy distribution (SED) as a continuum illuminating the absorber. Using the photoionization code
{\sc Titan}, recently we have shown that the absorption measure distribution (AMD) found for 
this object can be  successfully modelled as a
single slab of gas in total pressure (radiation+gas) equilibrium, contrary to the
usual models of constant density multiple slabs. We discuss the transmitted spectrum
that would be recorded by an observer after the radiation from the nucleus passes through the WA.
\end{abstract}

\section{Introduction}

The first ever evidence of X-ray absorption has been discussed by \citet{halpern1984}
who analysed the {\it EINSTEIN} data of QSO MR 2251-178. Since then, significant progress
has been made in the study of X-ray absorption by an ionized absorber present in the line of sight to the centers of active galaxies. This was
possible as a result of the advancement in the X-ray instrumentation which enabled the acquisition of high resolution X-ray data clearly showing numerous X-ray absorption lines in the spectra of $\sim$ 50\% of Seyfert 1 galaxies.
These observations facilitated the exploration of the physical properties as
well as the  ionization and thermal structure of the absorbing gas for many AGNs
\citep[and many others]{kaastra2002,rozanska2004,holczer2007,behar2009,detmers2011,stern2014}.

Absorption lines observed in the high resolution X-ray data together with
the photoionisation modelling can be used to quantify the strength of
absorption which is parametrized as absorption measure distribution (AMD) \citep{holczer2007};
i.e., the distribution of gas column density N$_{H}$ in ionization parameter $\xi$. There is no general consensus whether these
absorbers exist under pressure equilibrium or the number
density is constant inside them. Recently, \citet[hereafter AD15]{adhikari2015} have
shown that the warm absorber in Mrk 509 can be modelled with a single
plane parallel slab of gas in total (gas+radiation) pressure equilibrium. Using the
photoionization modelling with the constant pressure assumption, they were successful
in reproducing the observed AMD derived by \citet{detmers2011} from the 600 ks RGS
spectrum of {\it XMM-Newton}. 

In this paper, we focus on the transmitted spectrum that an observer would see if the WA is modelled as in AD15. We also discuss the future
prospects of fitting the modelled spectra to the high resolution data
that will be available from the advanced spectrometers in future X-ray missions; e.g. {\it Astro-H} and
{\it ATHENA}. This will allow us to constrain the physical parameters
of the absorbing cloud by fitting the observed absorption lines with relevant models.
We discuss the potential effect of the warm absorption on the relativistic Fe line around $6.4$ keV following the idea presented in \citet{rozanska2006}.

\section{Model parameters}

\label{sec:param}
We performed numerical simulations of the transmitted spectra with photoionization code
{\sc titan} \citep{dumont2000} using the shape of the SED of Mrk 509
constrained by \citet{kaastra2011} as the broad band continuum illuminating the WA. {\sc Titan} solves the full radiative
transfer equation and determines the physical state of the gas at
each depth assuming the local balance between ionization and
recombination of ions, excitations and de-excitations, local energy
balance and finally total energy balance. For the detail features of
the {\sc titan} code and its implementation in the study of WA, we refer the reader to the relevant papers \citep[]{collin2004,goncalves2006,rozanska2006}.
In this study, we assumed the accretion radiation is normally incident on plane
parallel slab of absorbing gas. Constant total
pressure assumption self-consistently allows the gas to be stratified
in density when it is illuminated by the incident accretion
radiation. The photoionization modelling requires the following
parameters: SED, number density of the gas, ionization degree
(all defined at the illuminated side of the WA) and the chemical
composition. We refer to the paper AD15 for the
definitions, equations, and values of the relevant parameters
used in our calculations.  
\begin{figure}
\label{fig:wa}
\centering
\includegraphics[width=7.9cm]{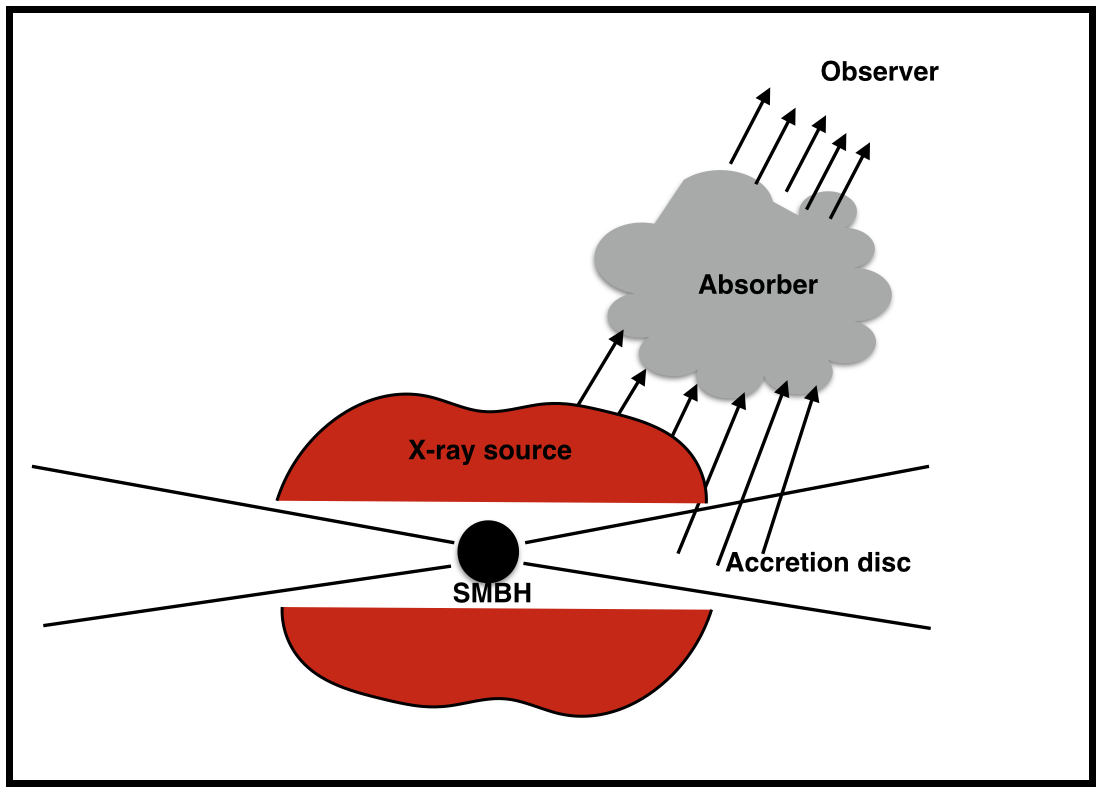}
\caption{ Schematic diagram showing the warm absorber. An observer sees absorption lines
in the radiation transmitted through the absorber.}
\end{figure}
%

\section{Results and Discussion}
\label{sec:results}
\begin{figure}
\centering
\includegraphics[width=13.3cm]{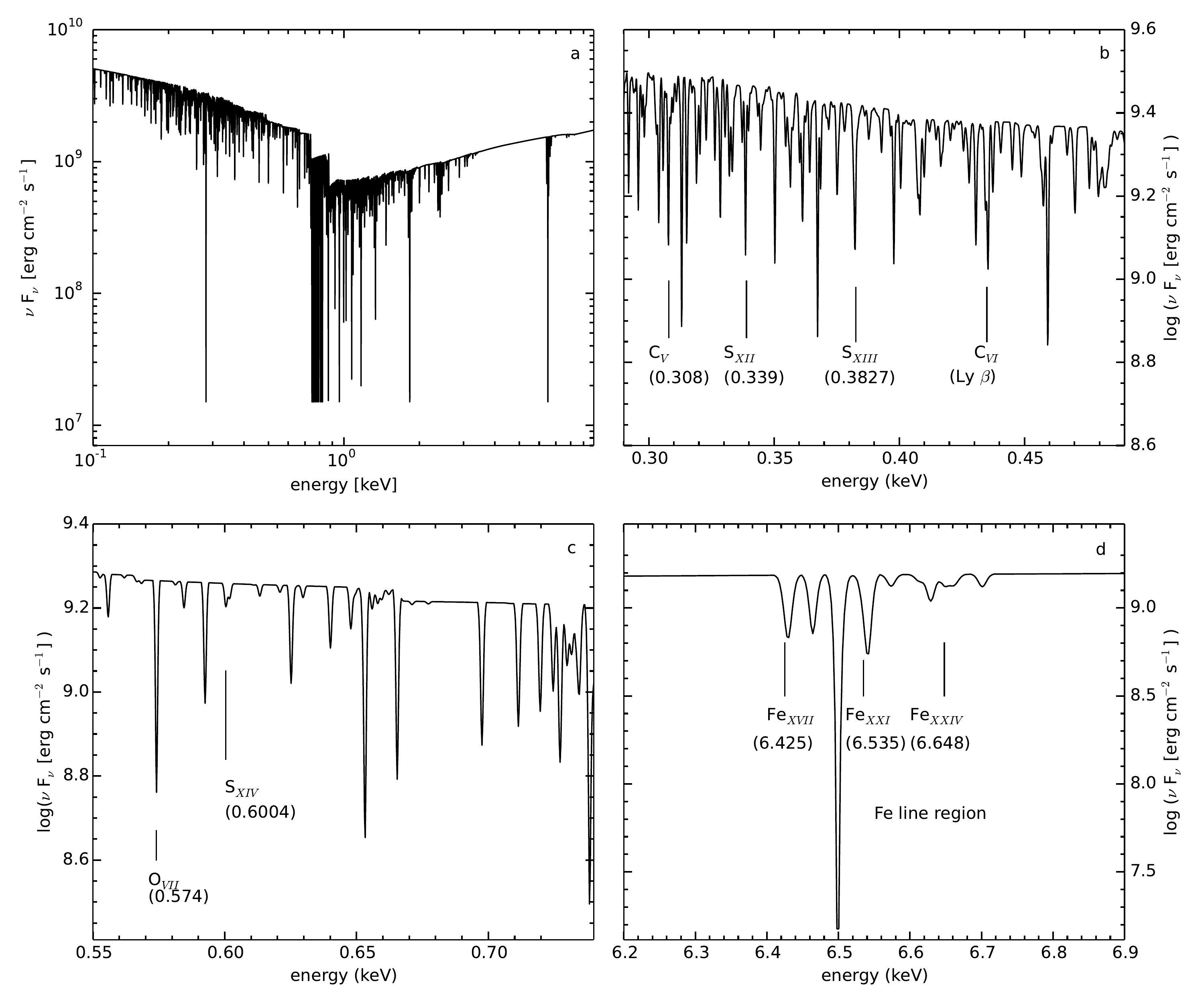}
\caption{Radiation spectrum transmitted through an absorber with the best fit parameters described in
 Fig. 8 of AD15.
Panel (a) shows the spectrum spanning the energy band
from $0.1-10$ keV. Panel (b) shows the transmitted spectrum zoomed in the energy range $0.3 -0.5$ keV. 
Panel (c) shows the spectrum zoomed in the energy range $0.55-0.75$ keV. In panel (d), the spectrum 
is zoomed in the Fe-line emission region around $6.4$ keV. Here the relativistically smeared reflection
 is not included.}
\label{fig:trans}
\end{figure}

The observed AMD found by \citet{detmers2011} can be succesfully modelled by the cloud 
compressed by 
radiation pressure as pressented in AD15 (see Fig.8). Absorption measure distribution in Mrk~509 exhibits 
two dipps which are reproduced by our photoionization model.
In this work, we discuss the continuum transmitted through the absorber described by the best fit 
parameters as given in AD15. Fig. \ref{fig:trans} (panel a) shows the transmitted spectrum from the 
best fitting WA model in Mrk~509. Numerous absorption lines
are expected in the spectrum when the radiation is absorbed by the ionized gas. The absorption lines 
produced due to ions of lower ionization degree are seen in the transmitted spectra 
 (panels (b) and (c) of Fig. \ref{fig:trans}) together with lines from higher ionization level from the one 
absrobtion component of the gas under constant pressure (panel d). 
Expanded view of the spectra in the Fe line region around 
$6.4$ keV is shown in panel (d). Around $6.4$ keV, we see a number of absorption lines due to highly
ionized Fe ions with significant equivalent widths. This result clearly shows that the WA has a potential
 to significantly affect the Fe line emission that may be present in the illuminating spectrum due to 
the reflection of the hard X-ray continuum on the disk surface. 
However, a quantitative conclusion can be drawn only after the confirmation of this result
in the spectra observed with high resolution instruments in the future X-ray missions, such as 
{\it Astro-H} and {\it ATHENA}.

\bibliographystyle{ptapap}
\bibliography{ptapapdoc}

\end{document}